\documentclass{article}

\usepackage{microtype}
\usepackage{graphicx} % Required for inserting images
\usepackage{subfigure}
\usepackage{booktabs} % for professional tables
\usepackage{amsmath}

\usepackage{tablefootnote}

\usepackage{float}

\usepackage{hyperref}
\usepackage{adjustbox}
\usepackage[most]{tcolorbox} % nice colored box
\usepackage{pgfplots}
\pgfplotsset{compat=1.18}
\usepackage{caption}
\usepackage{listings}
\usepackage[table]{xcolor}

\usepackage{pifont}
\usepackage[table]{xcolor}
\definecolor{TickGreen}{HTML}{DFF0D8}  % light green
\definecolor{LightGrey}{HTML}{F2F2F2}  % light grey
\newcommand{\cmark}{\ding{51}}       % tick mark
\newcommand{\greencell}{\cellcolor{TickGreen}{\cmark}}
\newcommand{\greycell}{\cellcolor{LightGrey}{\phantom{\cmark}}}
\usepackage[accepted]{icml2025}

\usepackage[capitalize,noabbrev]{cleveref}

\usepackage[textsize=tiny]{todonotes}
\newcommand{\mmore}{\textls[100]{\texttt{MMORE}}}
\definecolor{green1}{HTML}{5ACEA0}

\icmltitlerunning{MMORE: Massive Multimodal Open RAG \& Extraction
}

\begin{document}

\twocolumn[
\icmltitle{MMORE: Massive Multimodal Open RAG \& Extraction
}



\icmlsetsymbol{equal}{*}

\begin{icmlauthorlist}
\icmlauthor{Alexandre Sallinen}{epfl}
\icmlauthor{Stefan Krsteski}{epfl}
\icmlauthor{Paul Teiletche}{epfl}
\icmlauthor{Marc-Antoine Allard}{epfl}
\icmlauthor{Baptiste Lecoeur}{epfl}
\icmlauthor{Michael Zhang}{epfl}
\icmlauthor{David Kalajdzic}{epfl}
\icmlauthor{Matthias Meyer}{eth}
\icmlauthor{Fabrice Nemo}{epfl}
% \icmlauthor{Charlotte Meyer}{epfl}
% \icmlauthor{Matthew Meyer}{epfl}
% \icmlauthor{Lea Grieder}{efpl}
% \icmlauthor{Laetitia Wilhelm}{eth}
\icmlauthor{Mary-Anne Hartley}{epfl,harv}
\end{icmlauthorlist}

\icmlaffiliation{epfl}{École Polytechnique Fédérale de Lausanne (EPFL), Switzerland}
\icmlaffiliation{eth}{ETH Zürich, Switzerland}
\icmlaffiliation{harv}{T.H. Chan School of Public Health, Harvard University, USA}

% You need this line to close the twocolumn block and make affiliations appear:

\centering
\begin{tabular}{@{}c@{\hspace{0.5cm}}c@{}}
$^1$ EPFL, Switzerland & $^2$ ETHZ, Switzerland \\
\multicolumn{2}{c}{$^3$ Harvard University, USA} \\
\end{tabular}
\vskip 0.3in
]
\begin{figure*}[t]
    \centering    \includegraphics[width=0.75\linewidth]{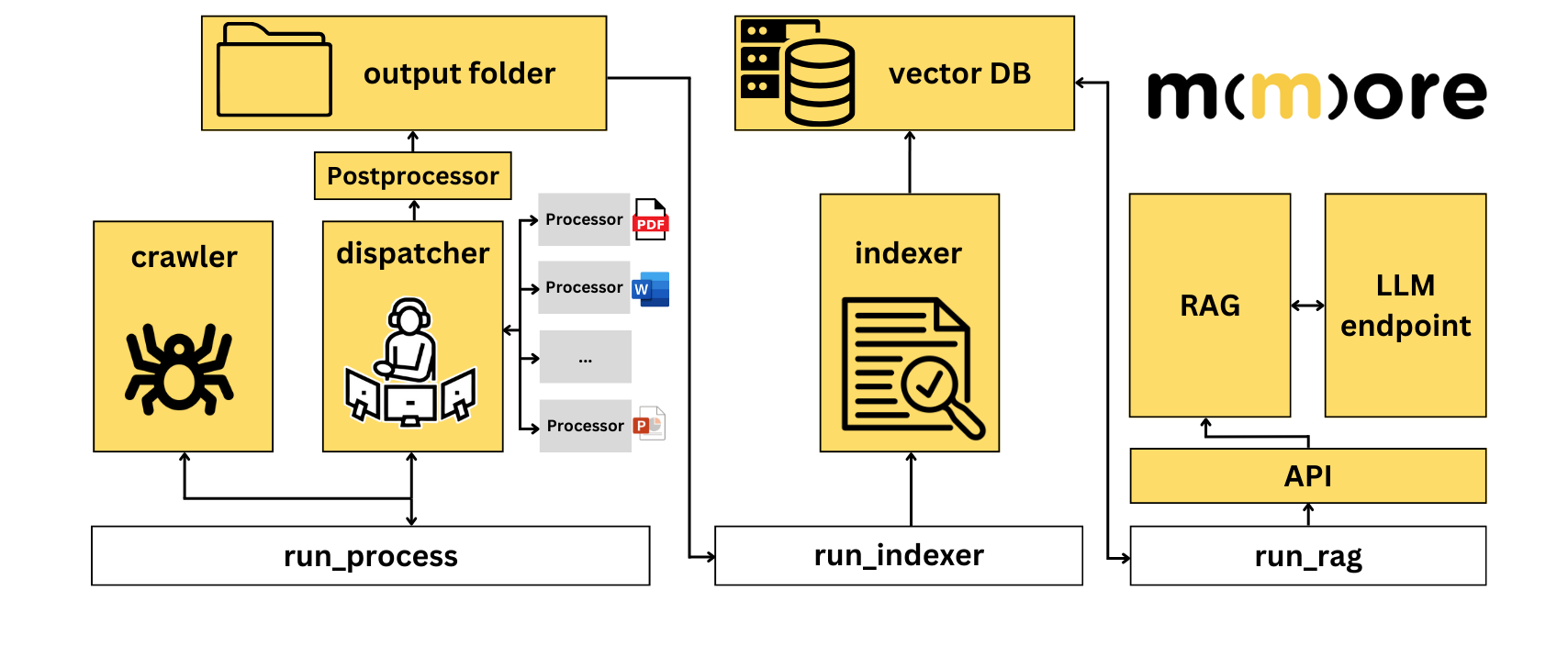}
    \vspace{-3mm}
    \caption{The end-to-end pipeline from file-type–specific processing to retrieval-augmented generation (RAG).}
    \label{fig:mmore_system_overview}
\end{figure*}

\begin{abstract}
We introduce \textbf{\mmore{}}, an open-source pipeline for \textbf{\texttt{M}}assive \textbf{\texttt{M}}ultimodal \textbf{\texttt{O}}pen \textbf{\texttt{R}}etrieval-Augmented Generation and \textbf{\texttt{E}}xtraction, designed to ingest, transform, and retrieve knowledge from heterogeneous document formats at scale. \mmore{} supports more than fifteen file types, including text, tables, images, emails, audio, and video, and processes them into a unified format to enable downstream applications for LLMs. The architecture offers modular, distributed processing, enabling scalable parallelization across CPUs and GPUs. On processing benchmarks, \mmore{} demonstrates a 3.8-fold speedup over single-node baselines and 40\% higher accuracy than Docling on scanned PDFs. The pipeline integrates hybrid dense-sparse retrieval and supports both interactive APIs and batch RAG endpoints. Evaluated on PubMedQA, \mmore{}-augmented medical LLMs improve biomedical QA accuracy with increasing retrieval depth. \mmore{} provides a robust, extensible foundation for deploying task-agnostic RAG systems on diverse, real-world multimodal data. The codebase is available at \href{https://github.com/swiss-ai/mmore}{https://github.com/swiss-ai/mmore}.
    
\end{abstract}

\section{Introduction}
As of 2025, the public web is conservatively estimated to host more than 2.5\,trillion PDF documents, alongside petabytes of mixed-modality slide decks, spreadsheets, images, and audiovisual artefacts \citep{pdfscloudfiles}.  Yet fewer than one percent of these resources are represented in popular machine-learning corpora as they are remain locked behind brittle, heterogeneous formats that frustrate automated parsing at scale. Existing pipelines rely on ad hoc mosaics of format-specific utilities, limiting throughput, reproducibility, and long-term maintainability.

As data-supply forecasts estimate that the pool of high-quality human-generated text could be exhausted by prevailing scaling trends as early as 2026 \cite{villalobos2022,llmdata2024}, it has become essential to find more format-agnostic preprocessing workflows. Much of this data, particularly in specialized or institutional settings, is unavailable for training but remains crucial for improving the verifiability of LLM outputs through RAG. Hallucinations \citep{openai2025o3o4} and factual drift \citep{huang2025survey} remain significant challenges, and robust RAG pipelines are increasingly explored as a means to mitigate these issues, thereby reducing the burden of manual validation and better aligning model outputs with trustworthy source material.

To address these limitations, we introduce \mmore{} an open-source tool for \textbf{\texttt{M}}assive \textbf{\texttt{M}}ultimodal \textbf{\texttt{O}}pen \textbf{\texttt{R}}etrieval-Augmented Generation and \textbf{\texttt{E}}xtraction, a unified pipeline for scalable extraction, transformation, and retrieval of multimodal data. \mmore{} supports diverse formats such as documents, presentations, spreadsheets, and multimedia and integrates them into a structured knowledge base, enabling LLMs to access accurate, contextually grounded information via the RAG paradigm.

Designed for modularity and scalability, our pipeline natively supports parallelized processing across multi-node architectures and distributed environments such as Kubernetes clusters. Compared to Docling demonstrates more than 2-fold faster end-to-end processing, while achieving 40\% higher layout accuracy on scanned PDFs. In distributed mode, we show that our pipeline processes 720 pages in 185s using four nodes, resulting in 3.8-fold speedup over single-node mode. The results demonstrate \mmore{}'s effectiveness as a scalable, high-accuracy solution for multimodal document processing in real-world deployment.

% MMORE runs seamlessly on a single machine or scales up to large compute clusters, fully utilizing available resources. It can efficiently handle terabytes of data — think 8000 PDFs, 2000 videos, and 500 spreadsheets — and prepare them for RAG-based systems.

% Designed with software engineering best practices, MMORE prioritizes modularity, interoperability, and ease of extension. It builds on reliable, actively maintained tools such as Dask (distributed compute), FastAPI (API layer), and PyTorch (ML workflows), and integrates natively with LangChain, Hugging Face, and Milvus to support seamless RAG pipeline construction. This foundation ensures high performance and long-term sustainability.

% \begin{tcolorbox}[colback=blue!5!white]
% \small
% \textbf{MMORE Design Philosophy} \\
% 1. \textbf{Modularity} — Plug-and-play components \\
% 2. \textbf{Interoperability} — Integrates with LangChain, Hugging Face, Milvus \\
% 3. \textbf{Extensibility} — Add formats or models quickly \\
% 4. \textbf{Sustainability} — Built on trusted, maintained tools
% \end{tcolorbox}
\section{Related Work}
% % mention LLMWhisperer (this is not open-source and its paid, so our contribution here is easy)
% % doctr - they only do document parsing 
% % Surya - this one is the most relevant because we use it in our pipeline for PDFs intesively. The difference is we only reuse surya for PDFs and we offer native parallelization on multi-node multi-gpu systems in contrast

Large-scale transformation of unstructured documents into structured, machine‑readable format has attracted substantial attention. We group prior work into two strands: \textbf{(i)} document ingestion and parsing pipelines, and \textbf{(ii)} RAG frameworks. To our knowledge, neither line of work simultaneously offers the modality coverage and end‑to‑end throughput required for industrial‑ and small‑scale multimodal assistants that we target with \mmore{}.\\
\textbf{Document Ingestion Pipelines.} GPU‑accelerated microservice suites such as \textit{NV‑Ingest}~\cite{nvingest} convert PDFs and office documents into page‑level JSON enriched with text blocks, tables, and graphics, and can optionally export embeddings for downstream indexing. \textit{Docling}~\cite{auer2024docling} extends the modality set to spreadsheets, and other common formats, but executes primarily on a single node and therefore exhibits limited throughput in production settings. Classical OCR tools like \textit{doctr}~\cite{doctr2021} handle text detection and recognition but rely on external systems for layout, embeddings, and indexing. \textit{Surya}~\cite{paruchuri2025surya} adds multilingual OCR and layout analysis but lacks built-in multi-GPU or cluster parallelism. Commercial services such as \textit{LLMWhisperer}~\cite{llmwhisperer2025} offer similar functionality behind a paywall, which restricts reproducibility and hinders open experimentation. In contrast, \mmore{} combines extraction, transformation, embedding, and indexing into a single open‑source pipeline that natively parallelizes across multi‑node, multi‑GPU deployments. Moreover, \mmore{} uniquely handles audiovisual assets, enabling unified RAG over text, images, and time‑based media. \\
\textbf{RAG Frameworks.} Open‑source libraries such as \textit{LangChain}~\cite{Chase_LangChain_2022} and \textit{LlamaIndex}~\cite{Liu_LlamaIndex_2022} provide high-level abstractions for chunking, embedding, retrieval, and prompting. However, they rely on external loaders for modality‑specific parsing and give no guidance on efficient high-throughput ingestion. 
Several recent pipelines, such as \textit{Unstructured.io}~\cite{unstructured2025} and Haystack~\cite{haystack2019} for document parsing, or \textit{M3IT}~\cite{li2023m3it} and \textit{OpenFlamingo}~\cite{awadalla2023openflamingo} for multimodal model alignment, address specific components of this pipeline. Yet none provide an integrated, open-source framework that supports ingestion, transformation, and retrieval across heterogeneous, real-world file types at scale. 

\mmore{} combines a scalable ingestion layer with a task-agnostic retrieval API, unifying document processing and RAG tools to enable multimodal assistants from raw enterprise data in one library.
%%% vérifier les sourches encore

\section{Architecture}

% MMORE provides an end-to-end RAG platform: you can use the same tool for processing your database of documents, making an index out of it, and query the LLM of your choice with the relevant documents.
\mmore{} provides an end-to-end platform, enabling users to process large document collections, build retrieval indices, and query LLMs with relevant multimodal content, all within a unified framework, as illustrated in Figure~\ref{fig:mmore_system_overview}.

\subsection{Processing}

At the core of \mmore{} lies a modular, scalable processing pipeline, designed for efficient, multimodal data extraction. Importantly, \mmore{} reuses open-source extraction tools such as \textit{Surya}~\cite{paruchuri2025surya} for PDF parsing, \textit{Whisper}~\cite{radford2023robust} for audio transcription, and standard Python libraries for office file formats, allowing us to focus on scalable orchestration and integration. A complete list of supported extractors is provided in Appendix~\ref{app:ingestion}. The design prioritizes three main strengths: \textbf{(i)} multimodal document processing, \textbf{(ii)} extensibility to new file types, and \textbf{(iii)} high-throughput distributed execution.\\
\textbf{Multimodal Data Extraction.}
\label{processing:multimodal}
The processor module extracts heterogeneous content from documents and standardizes it into a unified JSON-based format, referred to as the \textit{MultimodalSample} (see Appendix \ref{app:data_format}). Each sample consists of plain text interleaved with modality placeholders (e.g. images) and a list of the extracted modalities, preserving their type and location. Embedded media are extracted and saved to disk, with placeholder tokens (e.g., \texttt{<attachment>}) inserted at the corresponding positions within the text. This design supports downstream tasks that require text with tightly linked visual elements, such as multimodal pre-training or RAG.\\
\textbf{Extensibility.}
\label{processing:extensibility}
To facilitate extensibility, we designed a common processor interface that abstracts file-specific handling into modular components. Adding support for a new file type requires only implementing a lightweight subclass, promoting long-term maintainability and community-driven contributions. Each processor needs to define a class that takes a file path as input and outputs a \textit{MultimodalSample}, leveraging the standardized output format across the system. To date, \mmore{} supports more than 15 file types, including, but not limited to, PDFs, DOCX, PPTX, spreadsheets, media files, emails, and HTML pages.\\
\textbf{Distributed Processing.}
\label{processing:distribution}\mmore{} natively supports both intra-node and inter-node parallelization, exploiting all available CPU and GPU resources without requiring manual configuration from the user. The system is built on top of \textit{Dask}~\cite{dask}, enabling automatic workload balancing, fault tolerance, and seamless scaling across deployment settings, from standalone machines to large multi-node clusters. This design scales across use cases, from individual researchers to large organizations. To further support both ends of the spectrum, \mmore{} offers two processing modes: a fast mode for speed and a default mode for accuracy, allowing users to balance performance and fidelity as needed.

% at the end 
% To accommodate different computational needs, we provide both \textit{fast} and \textit{slow} processing modes, offering a trade-off between higher accuracy with slower processing and lower accuracy with faster processing. 

\subsection{RAG}

The RAG pipeline is composed of three independent components: \textbf{(i)} post-processing, \textbf{(ii)} indexing and retrieval, and \textbf{(iii)} an integrated RAG service. Each part is modular and can be run independently.

% old version Those parts can be run independently, allowing users to perform only the operation they want. They also each have a configuration file associated with them, allowing the user to easily fine-grain the settings depending on what the user wants.

\textbf{Post-processing.}
\label{RAG:postproc}
% old version Post-processing is meant to perform data processing on the extracted texts from documents. The infrastructure for post-processing is customizable; one can easily implement a new post-processor, and then the configuration for post-processing specifies which post-processors should be used and with which parameters. The post-processor is meant to produce a new JSONL file with the post-processed content.
This stage filters the extracted text to improve quality for downstream tasks. \mmore{} exploits the existing \textit{datatrove} \cite{penedo2024datatrove}, a high-throughput filtering library, and includes native support for several post-processing components, including Named Entity Recognition, Chunking, and Tagging. \\
% Additional post-processors can be integrated with minimal effort, and the entire pipeline is configured using lightweight YAML files. This stage produces a cleaned and optionally enriched JSONL dataset, ready for downstream indexing or training.
% old versio no.2 The infrastructure is modular and extensible: MMORE natively supports the following post-processors: Named Entity Recognition, Chunker, Tagger\footnote{The Tagger enhances the metadata of indexed documents based on their content}, and Filter\footnote{The Filter filters out documents that are deemed useless, based on predefined criteria}. New post-processors can easily be implemented, and pipelines can be configured through lightweight YAML files. The post-processing stage produces a new JSONL file containing cleaned and optionally enhanced document samples.
\textbf{Indexing and Retrieval.}
\label{RAG:indexing_retrieval}
% old version Indexing is a crucial component of the RAG pipeline, as the retrieval system has to deal with the data representation made by the indexer to retrieve the most relevant documents. We also made the indexing modular: MMORE currently supports two indexers, a regular RAG indexer based on dense and sparse embeddings computed for each document, and a GraphRAG. Users may define a new indexer.
Indexing is crucial to RAG performance, as retrieval relies on how documents are represented. \mmore{} uses a hybrid indexing strategy, storing both sparse and dense embeddings for each document. Sparse representations support lexical matching and improve interpretability, while dense embeddings enable semantic search using neural similarity. This duality allows users to choose or combine retrieval embeddings depending on their downstream task. The retriever is accessible via our integrated RAG system or as a standalone API.\\
\textbf{Integrated RAG system.}
\label{RAG:ragsystem}
The RAG system supports both API-based querying and offline batch processing. In batch mode, users provide a JSONL file containing retrieval queries; the system processes each entry and saves the results to a new JSONL file. Both modes allow customization of the model, prompt template, index source, and other parameters via configuration files or API options.

% The RAG system supports both API-based queries and as a batch-mode service. When using the latter, the user provides a JSONL file with requests, and the RAG pipeline processes each input, saving the responses in a new JSONL file. Users can customize the LLM, prompt template, index source, and other parameters in the configuration file (or through the API).
% \input{sections/pipeline}
\section{Evaluation Setup}
We evaluate \mmore{}'s processing and RAG modules independently. Below, we detail our methodology for assessing efficiency, accuracy, and scalability.

\subsection{Processing}
The processing module is evaluated along two axes: efficiency and accuracy, versus \textit{Docling}~\cite{auer2024docling} as a baseline due to its popularity and ease of use.

\textbf{Efficiency.} We benchmark processing speed using a single A100 80GB. For scalability analysis, we use an 18-page paper and synthetically generate longer documents by duplicating its content to reach 36, 54, 90, to 720 pages. This setup allows us to test throughput for both single-device and distributed processing. The distributed experiments are conducted on a Kubernetes cluster with 1 vs 4 nodes (1 A100 per node) to evaluate parallelization efficiency. To highlight \mmore{}'s strength in handling heterogeneous data, we also evaluate its performance across a diverse set of 19 files, spanning 9 unique file types.

\textbf{Accuracy.} To assess text extraction quality, we create a benchmark using public-domain books from Project Gutenberg~\cite{projectgutenberg} by pairing PDF inputs with their corresponding plain-text ground truths. We select two contrasting cases: "The Blue Castle" (a clean, digital-friendly PDF) and "The Great Gatsby" (a scanned, image-based file). Each document is truncated to 50k characters to ensure computational feasibility, 
particularly for metrics like Levenshtein distance. We report standard metrics: BLEU~\cite{bleuscore} for n-gram overlap,
ROUGE-L~\cite{lin2004rouge}, and character error rate (CER)~\cite{levenshtein}. Metric formulations are provided in the Appendix~\ref{app:metrics}

\subsection{RAG}
To evaluate our RAG pipeline, we focus on the PubMedQA benchmark~\cite{jin2019pubmedqa}, a biomedical question-answering task. We construct a retrieval corpus by indexing all PubMed abstracts and conclusions into a dense vector database using \mmore{}. At inference time, the top-\(k\) most relevant documents are retrieved using a similarity search and prepended to the original question as context for the language model. We experiment with both Meditron3-8B and Meditron3-70B~\cite{sallinen2025llama}, evaluating how different values of \(k\) affect downstream accuracy. This setup isolates the effect of retrieval depth on performance within a consistent biomedical knowledge source.
% \section{MMORE Performances \& Applications}
\section{Results}

\subsection{Processing}

% %%% Acc. table
\begin{table*}[t]
\centering
\small
\resizebox{0.72\textwidth}{!}{%
\begin{tabular}{lcccccc}
\toprule
 & \multicolumn{3}{c}{\textit{The Blue Castle (digital PDF)}} & \multicolumn{3}{c}{\textit{The Great Gatsby (scanned images)}} \\
\cmidrule(lr){2-4} \cmidrule(lr){5-7}
Method & BLEU\textbf{$\uparrow$} & ROUGE-L\textbf{$\uparrow$} & CER\textbf{$\downarrow$} & BLEU\textbf{$\uparrow$} & ROUGE-L\textbf{$\uparrow$} & CER\textbf{$\downarrow$} \\ 
\midrule
\rowcolor{orange!10}
MMORE &0.8608 & 0.9940 & 0.0241 & \textbf{0.7973} & \textbf{0.9813} & \textbf{0.0295} \\
MMORE (fast)    & 0.8639 & \textbf{0.9963} & 0.0206 & 0.0000 & 0.0000 & 1.0000 \\
Docling         & \textbf{0.8643} & 0.9959 & \textbf{0.0199} & 0.5451 & 0.6582 & 0.5518 \\
\bottomrule
\end{tabular}%
}
\caption{Accuracy evaluation on two Project Gutenberg books: ``\textit{The Blue Castle}'' and ``\textit{The Great Gatsby}''.}
\label{tab:accuracy_results}
\end{table*}

%% Eff. figure
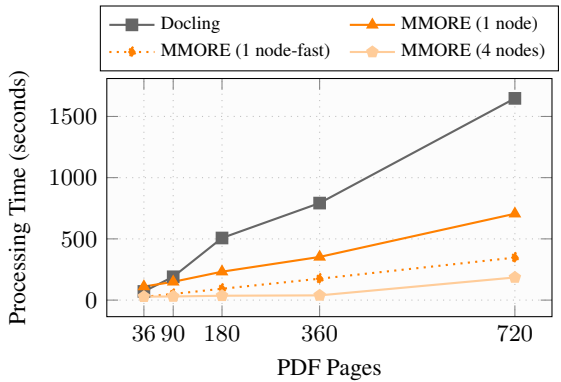
\begin{figure}[ht]
    \centering
    \begin{tikzpicture}
        \begin{axis}[
            xlabel={PDF Pages},
            ylabel={Processing Time (seconds)},
            legend style={
                font=\scriptsize,
                cells={anchor=west},
                at={(0.5,1.03)},
                anchor=south,
                legend columns=2,
                /tikz/every even column/.append style={column sep=0.5em}
            },
            grid=both,
            grid style={dotted,gray!50},
            xtick={36,90,180,360,720},
            width=7.5cm,
            height=4.75cm,
            scaled ticks=false,
            ticklabel style={/pgf/number format/1000 sep={}},
            label style={font=\footnotesize},
            tick label style={font=\footnotesize},
            every axis plot/.append style={thick},
            axis background/.style={fill=gray!2},
        ]

        % Docling (Black)
        % \addplot[color=black!50, mark=square*] table {
        %     36 70.66
        %     90 190.44
        %     180 507.23
        %     360 792.43
        %     720 1647.56
        % };
        % \addlegendentry{Docling}

        % Docling (Blue)
        \addplot[color=black!60, mark=square*] table {
            36 70.66
            90 190.44
            180 507.23
            360 792.43
            720 1647.56
        };
        \addlegendentry{Docling}
        
        % MMORE (1 node) - orange solid
        \addplot[color=orange, mark=triangle*, solid] table {
            36 110.67
            90 149.35
            180 232.05
            360 352.53
            720 705.33
        };
        \addlegendentry{MMORE (1 node)}

        % MMORE (1 node-fast) - orange dotted
        \addplot[color=orange, mark=diamond*, dotted] table {
            36 26.78
            90 50.29
            180 92.59
            360 174.66
            720 347.09
        };
        \addlegendentry{MMORE (1 node-fast)}

        % MMORE (4 nodes) - light orange solid
        \addplot[color=orange!40, mark=pentagon*, solid] table {
            36 28.19
            90 29.78
            180 35.71
            360 38.64
            720 185
        };
        \addlegendentry{MMORE (4 nodes)}

        \end{axis}
    \end{tikzpicture}
    \caption{Processing time vs. PDF length for \textit{Docling} and \mmore{}. \mmore{} (1 node-fast) disables OCR for performance, and \mmore{} (4 nodes) uses distributed processing.}
    \label{fig:pdf_processing}
\end{figure}

\textbf{Efficiency.} Figure~\ref{fig:pdf_processing} shows comparison of \textit{Docling} and \mmore{}. On short documents (36 pages) \textit{Docling} is marginally faster than \mmore{} (default). The difference disappears at 90 pages and shifts in favor of \mmore{} beyond 180 pages, where our pipeline scales almost linearly while \textit{Docling} slows down super-linearly. The fast mode, which omits OCR, delivers an additional speed-up of roughly two to three times. Running the default pipeline on four nodes achieves a 3.8-fold reduction in latency compared to the single-node baseline, surpassing even the single-node fast mode and clearly demonstrating the efficiency and scalability of the distributed execution in \mmore{}. It is also worth mentioning that the batch size is user-configurable. The experiments presented here used a conservative default, leaving around 65GB of the 80GB GPU unused. This highlights the potential for further optimization, as users can adjust the configuration to fully exploit available hardware resources. Table~\ref{tab:general_performance} further illustrates the performance advantage of \mmore{} across multiple file types. In default mode, \mmore{} reduces the total processing time by 45.48\% compared to \textit{Docling}, with the fast mode achieving an even more pronounced improvement of 155.38\%.

\begin{table}[ht]
\centering
\resizebox{0.46\textwidth}{!}{%
\begin{tabular}{@{}cccc@{}}
\toprule
\textbf{Metric} & \textbf{Docling} & \textbf{MMORE default} & \textbf{MMORE fast} \\
\midrule
Total Time (s)      & 522.98 & 358.93 & 204.57 \\
Num. of Unsupported Files    & 5 & 0 & 0 \\
Relative Efficiency & baseline & \textcolor{green1!90!black}{\textbf{+45.48\%}} & \textcolor{green1!90!black}{\textbf{+155.38\%}} \\
\bottomrule
\end{tabular}%
}
\caption{Processing speeds for 9 unique file types - PDF, DOCX, EML, MD, MP4, MP3, PPTX, TXT, XLSX (19 files in total).}
\label{tab:general_performance}
\end{table}
\vspace{-0.2cm}

\textbf{Accuracy.} Table \ref{tab:accuracy_results} reports BLEU, ROUGE-L, and CER on two Project Gutenberg titles. On the digitally formatted "Blue Castle" book, all three systems achieve near-perfect scores, with \textit{Docling} attaining the lowest CER (1.99\%); however, differences remain negligible. The scanned version of "The Great Gatsby", an image-based document requiring OCR, provides a greater challenge. Here, \mmore{} fast predictably fails, as it omits OCR entirely. In contrast, \mmore{} default maintains high extraction fidelity, clearly outperforming \textit{Docling}, whose CER of 55\% indicates significant OCR errors. Although these results demonstrate the accuracy of our pipeline, further benchmarking on a larger and more diverse set of documents is necessary to robustly validate its generalization capabilities.

\subsection{RAG}

% \begin{table}[ht]
% \centering
% \begin{tabular}{@{}llcc@{}}
% \toprule
% \textbf{LLM} & \textbf{Mode} & \textbf{Retriever} & \textbf{PubMedQA} \\ 
%              &              &                     & $\pm$ 0.0193 \\ \midrule
% Meditron3-8B & No RAG & - & 0.754 \\ 
%              & With RAG (k=1) & bge-large-en-v1.5 & \underline{0.79} \\ 
%              & With RAG (k=3) & bge-large-en-v1.5 & \textbf{0.80} \\ \midrule

% Meditron3-70B & No RAG & - & 0.802 \\ 
%               & With RAG (k=1) & bge-large-en-v1.5 & \underline{0.81} \\ 
%               & With RAG (k=3) & bge-large-en-v1.5 & \textbf{0.82} \\ 
% \bottomrule
% \end{tabular}
% \caption{Results demonstrating the effectiveness of RAG in improving LLM performance on PubMedQA.}
% \label{tab:results_pubmed}
% \end{table}

\begin{figure}[h]
\centering
\begin{tikzpicture}
\begin{axis}[
    xlabel={$k$},
    ylabel={PubMedQA Acc. (\%)},
    xtick={0,1,3},
    ymin=74, ymax=83,
    legend pos=south east,
    grid=major,
    grid style={dotted,gray!50},
    width=7.5cm,
    height=4.75cm,
    mark size=3pt,
    scaled ticks=false,
    every axis plot/.append style={thick},
    axis background/.style={fill=gray!2},
    legend style={
            font=\footnotesize,
            cells={anchor=west},
        },
]

% Meditron3-8B line (orange shade, diamond marker)
\addplot[
    color=orange!70, 
    mark=diamond*
] coordinates {
    (0, 75.4)
    (1, 79.0)
    (3, 80.0)
};
\addlegendentry{Meditron3-8B}

% Meditron3-70B line (darker orange shade, pentagon marker)
\addplot[
    color=orange!90!black, mark=pentagon*
] coordinates {
    (0, 80.2)
    (1, 81.0)
    (3, 82.0)
};
\addlegendentry{Meditron3-70B}

\end{axis}
\end{tikzpicture}
\vspace{-0.3cm}
\caption{Effect of retrieved documents ($k$) on PubMedQA accuracy for Meditron models using \mmore{}'s built-in RAG.}
\label{fig:pubmedqa_rag}
\end{figure}
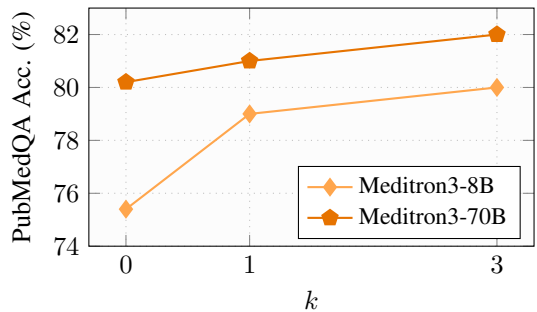
\vspace{-0.2cm}

To evaluate RAG performance, we test the Meditron-3 model family with various RAG configurations on the PubMedQA benchmark. Figure~\ref{fig:pubmedqa_rag} shows that both Meditron-3[8B] and Meditron-3[70B]~\cite{sallinen2025llama} consistently improve accuracy with RAG, especially as the number of retrieved documents $k$ increases. These results demonstrate that our RAG pipeline effectively injects domain-specific context at inference time, improving answer accuracy.

% RAG INFERENCE SPEED => GRAPH DB SIZE AS X, SPEED AS Y

% This demonstrate the effectiveness of using our RAG pipeline for enhancing the knowledge and performance of large language models (LLMs) in specialized domains, as exemplified by the PubMedQA benchmark. Our RAG system retrieves information from a vector database containing all PubMed abstracts and conclusions, ensuring access to relevant and high-quality context. Both Meditron3-8B and Meditron3-70B models \cite{meditron} show consistent improvements in accuracy when RAG is applied, particularly in settings where multiple retrieved documents are utilized. For instance, Meditron3-8B achieves a baseline accuracy of 0.7540 without RAG, which increases to 0.7900 with RAG using a single document and further improves to 0.8000 when three documents are retrieved. Similarly, Meditron3-70B exhibits a baseline accuracy of 0.8020, rising to 0.8100 with one document and reaching 0.8200 with three documents. These results highlight that our RAG pipeline not only retrieves the most relevant biomedical information effectively but also enables the LLM to leverage this context to improve its reasoning and accuracy. This confirms that our carefully designed RAG pipeline is highly effective when used in the right setting to maximize performance gains.
\section{Conclusion}
\mmore{} is a scalable, open-source pipeline for retrieval-augmented generation over diverse, real-world data. It supports more than 15 file types, including PDFs, spreadsheets, images, audio, and video, and enables structured, high-throughput integration into LLM workflows.

Our results show that \mmore{} outperforms \textit{Docling} in both speed and layout fidelity, particularly in OCR-heavy documents, and improves biomedical QA accuracy on PubMedQA via efficient RAG pipelines.

Built for extensibility and deployment at scale, \mmore{} provides a flexible foundation for verifiable, multimodal LLM applications. Future work will expand support for multilingual retrieval, audiovisual alignment, and federated processing in privacy-sensitive settings.

\bibliography{mmore}
\bibliographystyle{icml2025}

\onecolumn
\appendix
\clearpage
\section{Appendix}
\subsection{Document Ingestion}
\label{app:ingestion}
To better situate \mmore{} within the ecosystem of document ingestion systems, Table~\ref{tab:ingestion_comparison_expanded} presents a fine-grained comparison with two representative alternatives: \textit{Docling} and \textit{NV-Ingest} (part of NeMo Retriever). We evaluate them across modality support, indexing capabilities, and RAG integration. Green cells indicate native support, while grey cells denote the absence of the corresponding capability.

\begin{table}[h]
    \centering
    \small
    \resizebox{0.5\textwidth}{!}{%
    \begin{tabular}{lccc}
        \toprule
        \textbf{Feature} & \textbf{Docling} & \textbf{NV-Ingest\tablefootnote{\href{https://docs.nvidia.com/nemo/retriever/extraction/overview/}{NeMo Retriever Documentation}}} & \textbf{MMORE} \\
        \midrule
        % ---------------------- SUPPORTED MODALITIES ----------------------
        \rowcolor{gray!20}
        \multicolumn{4}{l}{\textit{Supported Modalities}} \\
        \midrule
        PDF & \greencell & \greencell & \greencell \\
        DOCX & \greencell & \greencell & \greencell \\
        PPTX & \greencell & \greencell & \greencell \\
        XLSX / spreadsheets & \greencell & \greycell & \greencell \\
        TXT & \greencell & \greencell & \greencell \\
        HTML & \greencell & \greycell & \greencell \\
        Markdown & \greencell & \greycell & \greencell \\
        CSV & \greencell & \greycell & \greencell \\
        Images (PNG/JPEG/SVG/TIFF/BMP) & \greencell & \greencell & \greencell \\
        Audio & \greycell & \greycell & \greencell \\
        Video & \greycell & \greycell & \greencell \\
        EML & \greycell & \greycell & \greencell \\
        \midrule
        % ---------------------- INDEXING & EMBEDDING ----------------------
        \rowcolor{gray!20}
        \multicolumn{4}{l}{\textit{Indexing \& Embedding}} \\
        \midrule
        Native engine included & \greycell & \greencell & \greencell \\
        LangChain / LlamaIndex connector & \greencell & \greencell & \greencell \\
        \midrule
        % ---------------------- RAG INTEGRATION ----------------------
        \rowcolor{gray!20}
        \multicolumn{4}{l}{\textit{RAG}} \\
        \midrule
        Built--in RAG pipeline & \greycell & \greycell & \greencell \\
        Plugin--based RAG & \greencell & \greencell & \greycell \\
        \midrule
        % ---------------------- LICENSE ----------------------
        Open--Source license & MIT & Apache~2.0 & Apache~2.0 \\
        \bottomrule
    \end{tabular}%
    }
    \caption{Fine-grained comparison of Docling, NV-Ingest, and MMORE document-ingestion pipelines. Green cells indicate native support; grey cells indicate absence of the capability.}
    \label{tab:ingestion_comparison_expanded}
\end{table}

\mmore{} supports a wide range of file formats through modular extractors. For each supported type, we define a \textit{default mode} prioritizing accuracy and a \textit{fast mode} optimized for speed. When no alternative tool is available, the fast mode is left unspecified (--). A complete list of tools used per file type is shown in Table \ref{tab:ingestion_tools}.

\begin{table*}[h]
\centering
\small
\resizebox{0.99\textwidth}{!}{%
\begin{tabular}{l l l}
\toprule
\textbf{File Type} & \textbf{Default Mode Tool(s)} & \textbf{Fast Mode Tool(s)} \\
\midrule
\rowcolor{gray!5}
DOCX & \texttt{python-docx} for text and image extraction & -- \\
MD & \texttt{markdown} for text, \texttt{markdownify} for HTML conversion & -- \\
\rowcolor{gray!5}
PPTX & \texttt{python-pptx} for text and image extraction & -- \\
XLSX & \texttt{openpyxl} for table and text extraction & -- \\
\rowcolor{gray!5}
TXT & Python built-in \texttt{open()} & -- \\
EML & Python built-in \texttt{email} module & -- \\
\rowcolor{gray!5}
Audio/Video (MP4, MP3, etc.) & \texttt{moviepy} for frames, \texttt{whisper-large-v3-turbo} for transcription & \texttt{whisper-tiny} \\
PDF & \texttt{marker-pdf} for OCR/structured data & \texttt{PyMuPDF} \\
\rowcolor{gray!5}
HTML & \texttt{BeautifulSoup} & -- \\
\bottomrule
\end{tabular}%
}
\caption{Overview of supported file types and extraction tools in \mmore{}. Full URLs are included in the project documentation.}
\label{tab:ingestion_tools}
\end{table*}

\subsection{Multimodal Sample}

The format provides a standardized representation for processed documents, combining extracted text with references to non-text elements. As shown in the example, the "text" field contains the document's content with \texttt{<attachment>} placeholders (which are configurable) marking modality locations, while the modalities array contains all embedded objects with their types and storage paths.

\label{app:data_format}
\begin{tcolorbox}[
  enhanced,
  colback=black!3,          % Light gray background inside
  colframe=black!50,        % Medium gray border
  coltitle=white,           % White title text
  fonttitle=\bfseries,      % Bold title font
  title=Format Example:, % Box title
  colbacktitle=black!50,    % Dark gray title background
  boxrule=0.4mm,
  toptitle=1mm,
  bottomtitle=1mm
]

\begin{lstlisting}[
    basicstyle=\small\ttfamily,
    numbers=none,
    backgroundcolor=\color{black!3}
]
{
  "text": "A report containing a cool image <attachment> and a chart <attachment>...",
  "modalities": [
    {
      "type": "image",
      "value": "chart_url_2.png"
    },
    {
      "type": "image",
      "value": "chart_url_1.png"
    }
  ]
}
\end{lstlisting}

\noindent\rule{\linewidth}{0.4pt}

\small{\textit{The standardized format for document processing.}}

\end{tcolorbox}

\subsection{Processing Accuracy - Metrics}
\label{app:metrics}

To quantify extraction accuracy, we used a combination of machine translation, summarization and string similarity metrics. Their definitions are given below.

\textbf{BLEU Score (bilingual evaluation understudy)} \cite{bleuscore}:  
The BLEU score evaluates the overlap between the n-grams (sequences of words of length \(n\)) between the extracted text and the ground truth. It is defined as:

\begin{equation}  
\text{BLEU} = \text{BP} \cdot \exp \left( \sum_{n=1}^{N} w_n \log p_n \right)
\end{equation}

where \(p_n\) is the precision for n-grams of length \(n\), ranging from [1 to 4], \(w_n\) are the weights (uniform), and brevity penalty (\(\text{BP}\)), given by:

\begin{equation}  
\text{BP} = 
\begin{cases} 
1 & \text{if } c > r \\ 
\exp\left(1 - \frac{r}{c}\right) & \text{if } c \leq r
\end{cases}
\end{equation}

Here, \(c\) is the length of the candidate (extracted) text, and \(r\) is the length of the reference (ground truth). BLEU considers how much of the extracted text matches the reference text in terms of word sequences, while also penalizing outputs that are too short.

\textbf{ROUGE-L (recall-oriented understudy for gisting evaluation)} \cite{lin2004rouge}:  
ROUGE-L measures the quality of the extracted text using the longest common subsequence (LCS) between the extracted text and the ground truth. The LCS is the longest sequence of words appearing in the same order in both texts (though not necessarily consecutively). ROUGE-L is calculated as:

\begin{equation}  
\text{ROUGE-L} = F_\text{measure} = \frac{(1 + \beta^2) \cdot \text{Precision} \cdot \text{Recall}}{\beta^2 \cdot \text{Precision} + \text{Recall}}
\end{equation}

where \(\beta\) is a weighting factor (set to 1 for equal weighting), and:

\begin{equation}  
\begin{aligned}
\text{Precision} &= \frac{\text{LCS}}{\text{Length of Extracted Text}}, \\
\text{Recall} &= \frac{\text{LCS}}{\text{Length of Ground Truth}}.
\end{aligned}
\end{equation}

\textbf{Levenshtein distance - character error rate (CER)} \cite{levenshtein}:  
Given two strings, \( s_1 \) (extracted text) and \( s_2 \) (ground truth), the Levenshtein distance \( d(s_1, s_2) \) measures the minimum number of insertions, deletions, or substitutions required to transform \( s_1 \) into \( s_2 \). We normalize this distance over the length of the ground truth and is defined as:

\begin{equation}
\text{CER} = \frac{d(s_1, s_2)}{|s_1|}
\end{equation}
\end{document}